\documentclass[sigconf]{acmart}
\usepackage{dirtytalk}
\usepackage{listings}

\copyrightyear{2022} 
\acmYear{2022} 
\setcopyright{acmlicensed}\acmConference[ITiCSE 2022]{Proceedings of the 27th ACM Conference on Innovation and Technology in Computer Science Education Vol 1}{July 8--13, 2022}{Dublin, Ireland}
\acmBooktitle{Proceedings of the 27th ACM Conference on Innovation and Technology in Computer Science Education Vol 1 (ITiCSE 2022), July 8--13, 2022, Dublin, Ireland}
\acmPrice{15.00}
\acmDOI{10.1145/3502718.3524795}
\acmISBN{978-1-4503-9201-3/22/07}

\begin{document}
\fancyhead{}

\title{XSS for the Masses: Integrating Security in a Web Programming Course using a Security Scanner}

\author{Lwin Khin Shar}
\orcid{0000-0001-5130-0407}
\affiliation{\institution{Singapore Management University}\country{Singapore}}
\email{lkshar@smu.edu.sg}

\author{Christopher M. Poskitt}
\orcid{0000-0002-9376-2471}
\affiliation{\institution{Singapore Management University}\country{Singapore}}
\email{cposkitt@smu.edu.sg}

\author{Kyong Jin Shim}
\orcid{0000-0002-1978-5384}
\affiliation{\institution{Singapore Management University}\country{Singapore}}
\email{kjshim@smu.edu.sg}

\author{Li Ying Leonard Wong}
\orcid{0000-0003-3250-3932}
\affiliation{\institution{Singapore Management University}\country{Singapore}}
\email{lywong.2017@smu.edu.sg}

\begin{abstract}
    Cybersecurity education is considered an important part of undergraduate computing curricula, but many institutions teach it only in dedicated courses or tracks.
    This optionality risks students graduating with limited exposure to secure coding practices that are expected in industry.
    An alternative approach is to integrate cybersecurity concepts across non-security courses, so as to expose students to the interplay between security and other sub-areas of computing.
    In this paper, we report on our experience of applying the \emph{security integration} approach to an undergraduate web programming course.
    In particular, we added a practical introduction to secure coding, which highlighted the OWASP Top 10 vulnerabilities by example, and demonstrated how to identify them using out-of-the-box security scanner tools (e.g.~ZAP).
    Furthermore, we incentivised students to utilise these tools in their own course projects by offering bonus marks.
    To assess the impact of this intervention, we scanned students' project code over the last three years, finding a reduction in the number of vulnerabilities.
    Finally, in focus groups and a survey, students shared that our intervention helped to raise awareness, but they also highlighted the importance of grading incentives and the need to teach security content earlier.
\end{abstract}



\maketitle

\section{Introduction}

Cybersecurity education is increasingly recognised as a critical component of undergraduate computing curricula.
This recognition is driven in part by the often staggering consequences of insecure code being exploited in the real world.
Accenture, for example, estimates the average annual cost of addressing and containing incidents to be $\$13$ million per firm, excluding any longer-term costs of remediation~\cite{McGinn19a}.
More recently, the remote code execution flaw in the widely-used Apache Log4j library has led to around 100 exploit attempts \emph{every minute} on software around the world~\cite{Corfield21a}.

Traditionally, cybersecurity has been taught as an elective topic through specialist courses, tracks (e.g.~\cite{Azadegan-03a}), or even dedicated degree programmes.
Unfortunately, this optionality means that a significant number of students may graduate with limited exposure to any of the secure coding practices that are demanded by industry.
This is because many of the core computing courses they take (e.g.~programming, software engineering) may not teach security at all, risking the perception that security is simply an afterthought, and potentially entrenching bad coding habits.
This problem has been discussed many times over the last two decades---as early as a SIGCSE'02 panel~\cite{Mullins-et_al02a}---but is yet to be sufficiently addressed.

A solution that has been advocated is the \emph{security integration} approach~\cite{Yue16a}, in which cybersecurity concepts are interweaved throughout non-security computing courses.
The approach ensures that every undergraduate---regardless of electives or track---reaches some baseline in cybersecurity knowledge.
But more importantly, it allows them to study cybersecurity in the context where it is needed, and thus demonstrates it as an indispensable part of different computing sub-areas.
In a database course, for example, this could be achieved by teaching SQL injection attacks as part of an introduction to relational database management systems~\cite{Taylor-Sakharkar19a}.
In a software engineering course, instructors could explore the consequences arising from a range of security anti-patterns, e.g.~social engineering attacks in systems with plaintext passwords.

In this paper, we report on our experience of applying the security integration approach to another computing sub-area: \emph{front-end web programming}.
This was motivated by the fact that insecurity is (regrettably) prevalent in client-side JavaScript across the web, e.g.~in the form of cross-site scripting~(XSS) vulnerabilities due to mishandled user input or the inclusion of outdated libraries~\cite{Lauinger-et_al17a}.
The course we targeted is part of our institution's core Information Systems curriculum, and since its inception, has focused on front-end fundamentals (i.e.~HTML, CSS, JavaScript, and frameworks) without any formal content on security.
In its latest iteration, we augmented the course with a practical 2-hour introduction to secure coding, which highlighted the OWASP Top 10 vulnerabilities~\cite{OWASP-Top10} using hands-on examples, and demonstrated how to identify them automatically using out-of-the-box security scanner tools such as ZAP~\cite{ZAP}.
Furthermore, we incentivised our students to utilise these tools in their group coding projects, offering bonus points to those who eliminated the most serious vulnerabilities from their work.

To evaluate the impact of this intervention, we scanned the code of 179 student projects over the last three years to determine the types of security vulnerabilities that students were introducing, as well as their prevalence and severity.
After explicitly introducing some security content to the course, we found that there were fewer vulnerabilities distributed across the projects.
Finally, we conducted a post-course survey and two focus groups, finding that our intervention helped to raise security awareness.
However, students also highlighted the importance of grading incentives in encouraging the adoption of secure coding, and the need to include the content as early as possible, i.e.~before releasing the project briefing.
Overall, we found security integration to be a promising, low-cost approach for raising general awareness of web security, as well as for allowing our most interested students to start developing their secure coding skills in context.

\section{Related Work}

The need for the security integration approach culminated in a SIGCSE'02 panel discussion, in which it was argued that the ``most effective way to incorporate security oriented issues into the curriculum is to include them as natural aspects of normal course topics''~\cite{Mullins-et_al02a}.
Perrone et al.~\cite{Perrone-Aburdene-Meng05a} called this the ``threading approach'', as it allows security and privacy to become a unifying theme of the overall curriculum, and helps avoid the isolation of knowledge units as may happen when security is only taught in dedicated courses.
Yue~\cite{Yue16a} revisited this argument more recently, arguing that security integration is still lacking in computing curricula, and that we are missing a ``golden chance'' to help ``students understand the correlation and interplay between cybersecurity and other sub-areas of [computing]''.
The motivation of our own intervention aligns with this point, in that we wanted to raise awareness of the impact of insecurity in the context of a web application.

Null~\cite{Null04a} observed that adding security topics to non-security courses can lead to some ``pleasant side-effects'', such as being great motivators for students, but that many instructors do not feel comfortable in their ability to teach them.
The author addressed this latter point by proposing concrete topics that raise security awareness without requiring deep proficiency.
Siraj et al.~\cite{Siraj-et_al14a} addressed it by creating an instructional support system intended to help non-security faculty by providing ready-made notes and assessment materials.
In our course, we addressed the issue by focusing on raising awareness (e.g.~the OWASP Top 10 examples) and by using automated security scanners (e.g.~ZAP) that can be run out-of-the-box and without expertise in the topic.

Several papers report on the need to integrate security into various non-security computing courses~\cite{Svabensky-et_al20a}.
Taylor and Kaza~\cite{Taylor-Kaza11a} augmented introductory programming courses with checklist-based `security injection modules', which students could follow to address issues such as buffer overflows and input validation.
Almansoori et al.~\cite{Almansoori-et_al20a} analysed student code from computer systems courses, finding widespread usage of unsafe C/C++ functions.
Similar to our approach, they recommend that instructors raise awareness by explaining the security implications of unsafe functions, and that secure coding be incentivised in exercises and projects using points.
In a subsequent paper~\cite{Almansoori-et_al21a}, it is highlighted that many computer systems textbooks do not address security, and do not teach how to use unsafe functions safely.
Taylor and Sakharkar~\cite{Taylor-Sakharkar19a} conducted a similar study but for undergraduate database textbooks with respect to one of the most common exploits: SQL injections.
While some textbooks went to the extent of providing in-depth coverage of defence, some had no mention of the topic at all and even provided SQL injectable code as examples of `correct' queries.
This is analogous to a problem we face in teaching web development, in that many online resources provide JavaScript code snippets that (if used improperly) could be exploited by XSS attacks.

A potential way to raise awareness of security concerns in different programming courses is to utilise IDE plug-ins that target insecure code.
Zhu et al.~\cite{Zhu-et_al13a} developed ASIDE, an Eclipse plug-in that provides security warnings for Java code, as well as explanations of vulnerabilities and some code generation.
The authors trialled the tool on 20 students, finding that it helped raise students' awareness when working on assignments.
Whitney et al.~\cite{Whitney-et_al15a} use a later version of the plug-in (now called ESIDE) in an advanced web programming course.
The authors observed that the tool helped students better understand why their code was insecure, but noted that it needed to be carefully incentivised to be effective, a point also emphasised by Tabassum et al.~\cite{Tabassum-et_al18a}.
In the aforementioned works, security warnings are provided to students `in the moment'~\cite{Whitney-et_al18a}, i.e.~while they are coding in the IDE.
Our approach of using security scanners (e.g.~OWASP ZAP) complements this by returning concrete (successful) attacks on students' code after subjecting it to state-of-the-art automated penetration tests.

\section{Context \& Intervention}

The context of our experience report is \emph{IS216: Web Application Development 2}, a core course taken in the second year of our undergraduate Information Systems degree programme.
IS216 introduces the fundamentals of front-end web programming, focusing on the HTML, CSS, and JavaScript languages.
Furthermore, it teaches the principles of responsive design using Bootstrap, as well as reactive user interfaces using the Vue.js framework.
IS216 follows on from previous courses that cover server-side programming and data management, which together give students the knowledge they need to develop full-stack web applications.

Our motivation for exploring the security integration approach arose on two fronts.
First, having taught the course for two years without any significant coverage of security concepts (other than one instructor including ad hoc tips without elaboration, e.g.~``using $\mathtt{eval}$ in JavaScript is bad''), we were concerned that students were unwittingly picking up programming habits that could lead to exploitable code in real-world projects.
We observed, for example, that students would frequently complete their JavaScript exercises using $\mathtt{innerHTML}$, a property that sets HTML content within an element on the page.
While extremely convenient, the property is susceptible to XSS attacks if it uses non-validated input (e.g.~injected HTML that the browser then renders).
We found that many students simply did not know that $\mathtt{innerHTML}$ was dangerous, aligning with Gollmann's observation that many security flaws in code arise mainly due to lack of awareness~\cite{Gollmann10a}.
Secondly, of all the courses in our degree, we felt that the security integration approach had the potential for most impact in a course on web development.
This is because front-end web development is all about \emph{user-facing code}, and instilling some basic security awareness could ensure that much of the low-hanging fruit, such as validating user inputs, gets `picked' by students.
Furthermore, while there are many online materials for learning web development, secure coding practices are often not mentioned.
For example, the W3Schools page for $\mathtt{innerHTML}$~\cite{W3Schools} does not mention the property's security implications at all.

Integrating security into IS216, however, required us to address a number of challenges.
First of all, previous runs of the course were already considered quite content-heavy, given that they required students to become adept at several different web languages/frameworks in a short period of time.
This raised the concern that too much additional content on security might make the course overwhelming for some students.
Second, as a large multi-section course, IS216 is taught by different faculty and instructors, most of whom are not security experts.
These challenges required us to design an intervention that raised awareness of web security while remaining accessible to students and teaching staff of all backgrounds.

\begin{figure}[!t]
\begin{center}
\includegraphics[width=0.9\linewidth]{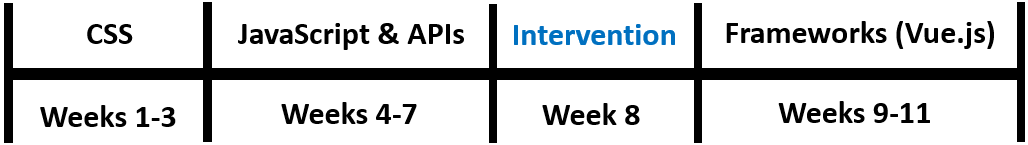}
\end{center}
\caption{IS216 lesson plan}
\label{fig:lessonplan}
\end{figure}

Hence, in this year's run of IS216, we decided to introduce students to security concepts in a more systematic way by offering an extra lecture in Week 8, our university's mid-term break period (Figure~\ref{fig:lessonplan}).
Furthermore, it was delivered to the \emph{entire cohort} as a single online class, rather than as separate lessons by the instructors teaching our 11 `sections'.
This ensured that: (1)~our established lesson plan was not significantly disrupted; and (2)~instructors less confident in security would not have to facilitate the session.

Our lecture was guided by the OWASP Top 10~\cite{OWASP-Top10} security vulnerabilities list.
In particular, we gave a practical, hands-on demonstration of the vulnerabilities by running the OWASP ZAP~\cite{ZAP} security scanner, step-by-step, on multiple relatable examples.
The idea was to make security more `visible' to students by showing them what could happen in practice (rather than simply in theory), and by relating our demonstrations to the JavaScript methods and properties that they were taught in previous weeks.

In order to drive interest in this additional break-week lecture, instructors factored students' attendance towards their class participation grades (i.e.~a holistically awarded grade component for fully participating in IS216 classes over the term).
Furthermore, to incentivise students into applying what they learnt, bonus marks were given to those who incorporated secure coding practices into their group projects.
Students could demonstrate this by highlighting their good practices (e.g.~validating user input) or by showing us good reports after running security scanners such as ZAP on their code.
As we were trialling this intervention for the first time, we offered only a modest number of bonus marks (i.e.~2 marks out of 25) for the project to mitigate the risk of teams diverging too far from the main learning objectives of the course.

\section{Scanning Student Projects}

Using ZAP~\cite{ZAP}, we scanned 87 projects from the previous two runs of IS216, as well as 92 projects from this year's run. 
To understand what could happen if the students really make an effort to improve security, we will also discuss the scan results of the projects for which the students explicitly declared that they are putting considerable effort into security.
There were 15 such projects amongst the 92 from this year's run.
Table~\ref{tab:sum} shows a summary of the results.

First, we must highlight that amongst this year's projects, there is an outlier project which reported 2,569 vulnerabilities in total.
As shown in Table~\ref{tab:sum}, this outlier causes a very high standard deviation of the results for this year's run (268.39 compared to 65.12 in the previous two runs and 63.5 in this year's run without the outlier).
Therefore, to avoid misinterpreting the results, in the following discussions we focus on comparing the results of this year's run without the outlier project against results of the last two runs.

\begin{table}[!t]
\caption{Summary of the results}
\label{tab:sum}
	\centering\footnotesize\vspace{-3pt}
	\begin{tabular}{lcccc}
		& prev.~2 runs & this year & this year & security \\ 
		& & w.~outlier & w.o outlier & focused \\ \hline
		No.~of projects & 87 & 92 & 91 & 15 \\
		Mean \#vuln & 72.74 & 94.23 & 67 & 68.33 \\
		Std Dev \#vuln & 65.12 & 268.39 & 63.5 & 103.85 \\
		Median \#vuln & 56 & 51.5 & 50 & 34
	\end{tabular}
\end{table}

\begin{figure}[!t]
\begin{center}
\includegraphics[scale=0.35]{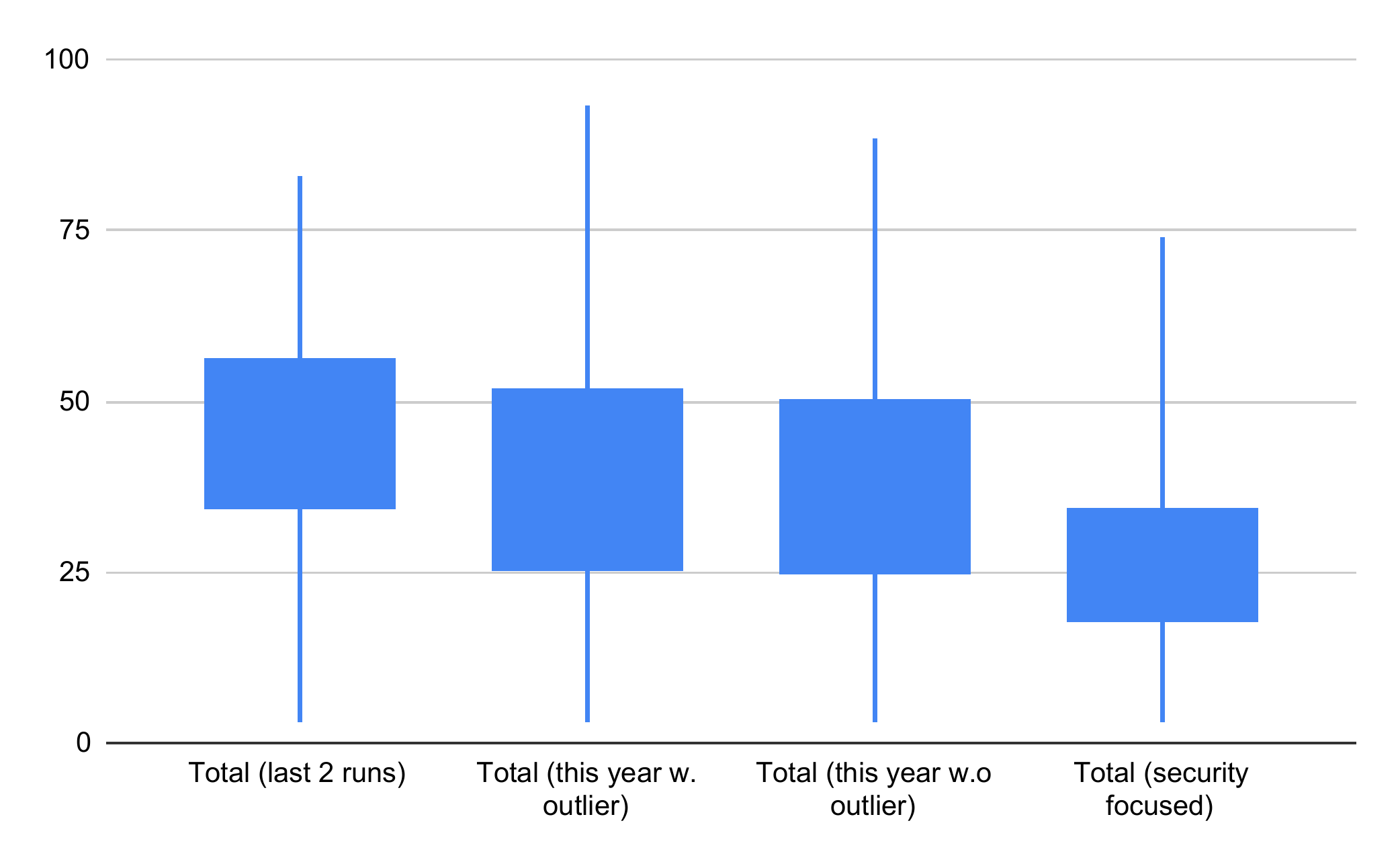}
\end{center}
\vspace{-1.5em}
\caption{Distributions of total vulnerabilities in last two runs' projects vs.~this year's projects (with and without the outlier) and this year's projects that focused on security}
\label{fig:totalvuln}
\vspace{-0.5em}
\end{figure}

As shown in Table~\ref{tab:sum}, the mean number of vulnerabilities in the projects from the last two runs is 72.74 and the mean number of vulnerabilities in this year's projects without the outlier is 67. 
The distribution of the scan results is shown in Figure~\ref{fig:totalvuln}. The leftmost boxplot in the figure shows the distribution of the total vulnerabilities found in the projects from the last two runs; the second boxplot shows the distribution of the total vulnerabilities found in this year's projects; the third plot shows the distribution of total vulnerabilities among this year's projects without the outlier; and the rightmost boxplot shows the distribution of total vulnerabilities among this year's projects that declared an emphasis on security.
As shown in Figure~\ref{fig:totalvuln}, the distribution of total vulnerabilities dropped in this year's run compared to the previous two runs.

\begin{figure}[!t]
\begin{center}
\includegraphics[scale=0.35]{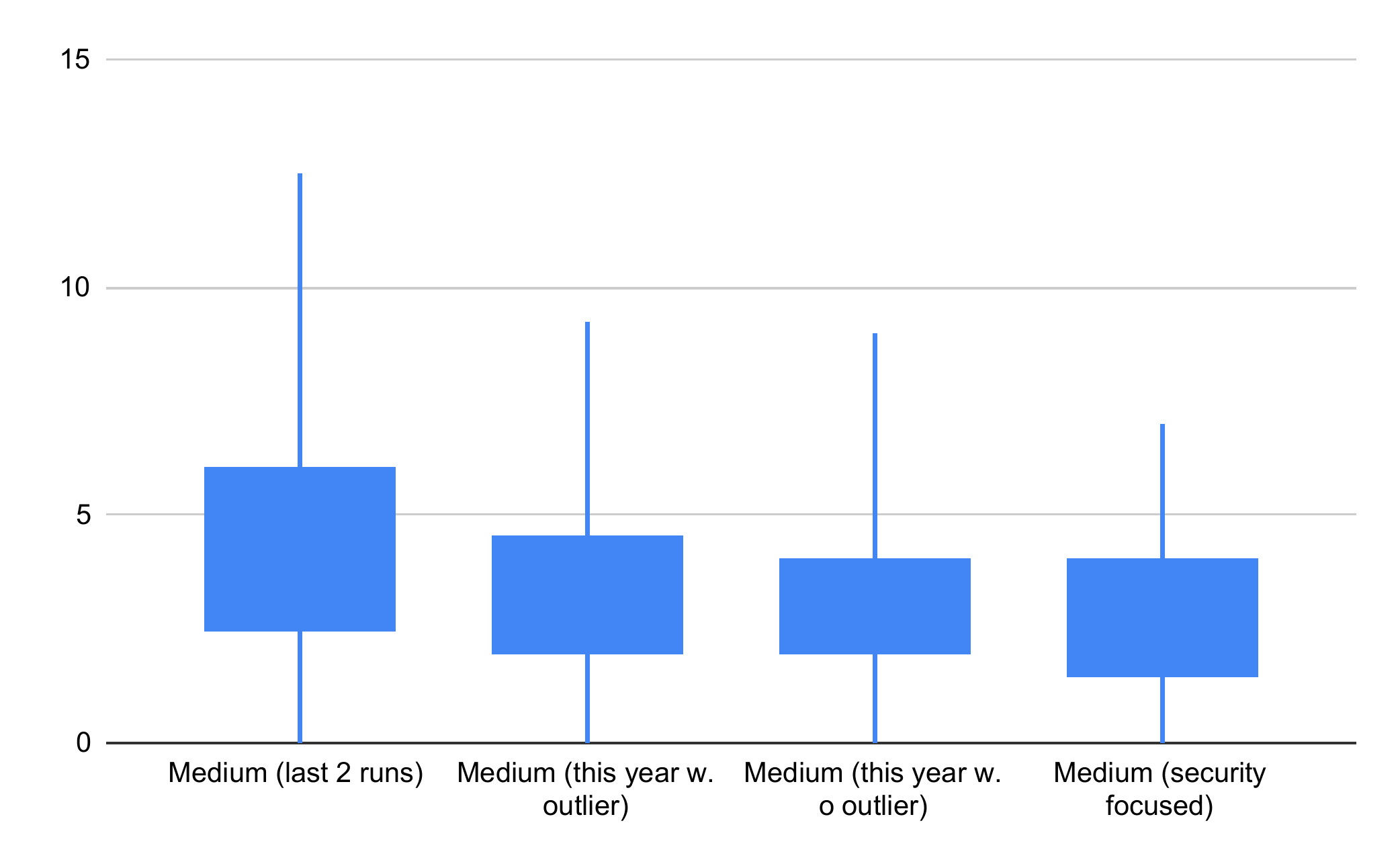}
\end{center}
\vspace{-1.5em}
\caption{Medium risk vulnerabilities found in this year's projects vs.~past years' vs.~projects that focused on security}
\label{fig:medium}
\vspace{-1em}
\end{figure}

\begin{figure}[!t]
\begin{center}
\includegraphics[scale=0.35]{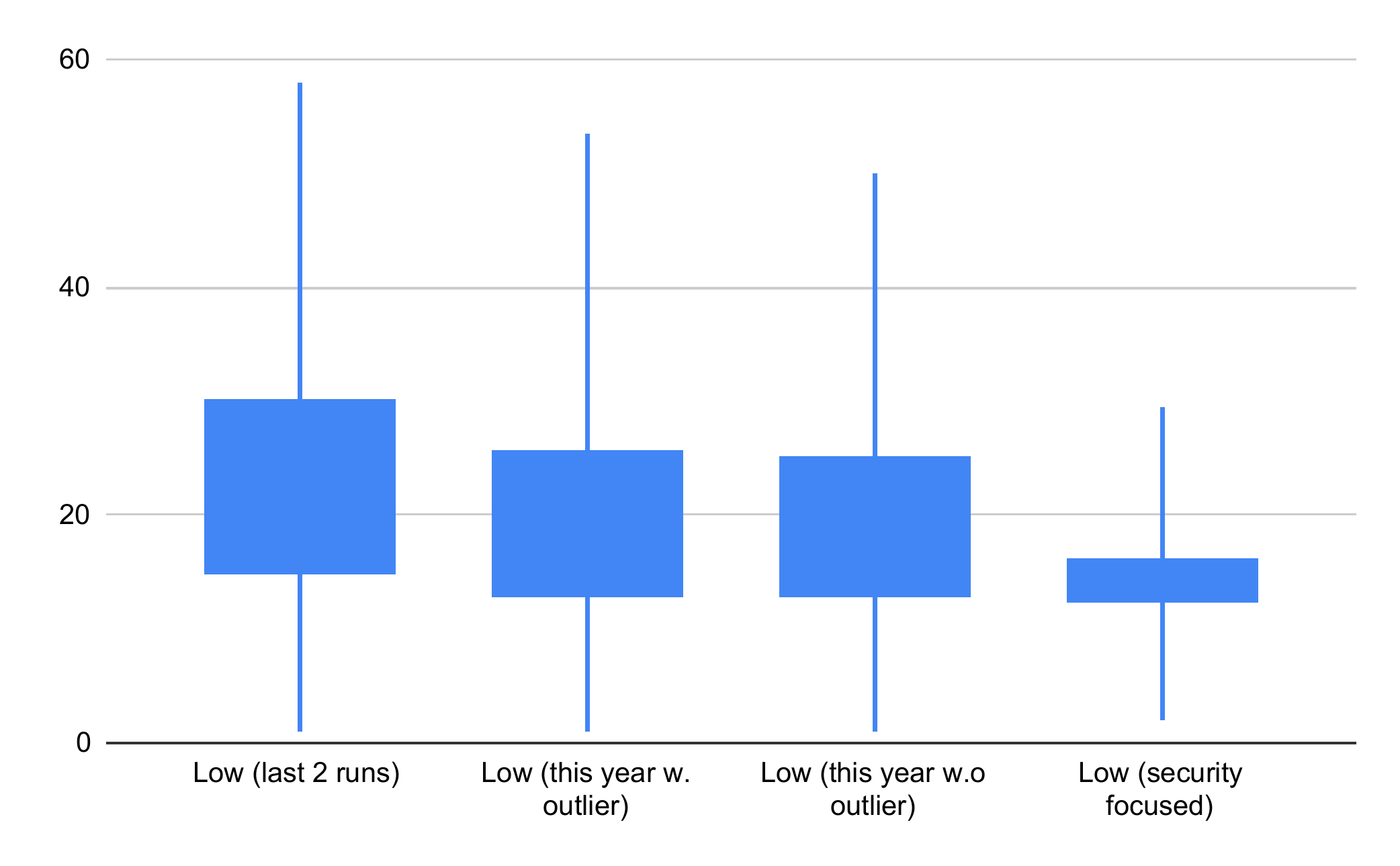}
\end{center}
\vspace{-1.5em}
\caption{Low risk vulnerabilities found in this year's projects vs.~past years' vs.~projects that focused on security}
\label{fig:low}
\vspace{-1em}
\end{figure}

In addition to the total vulnerabilities, ZAP also reports their risk level (severity).
Table~\ref{tab:risk} shows the percentages of the different risk levels of the vulnerabilities.
If we do not consider the outlier, the percentages of the medium and low risk vulnerabilities dropped in this year's projects compared to the previous two runs' projects.
This is further illustrated by the boxplots in Figures~\ref{fig:medium} and~\ref{fig:low}.

On the other hand, we observed that the distributions of high risk and informational risk vulnerabilities among the previous two runs' projects and this year's projects are similar (due to space constraints, we do not show the boxplots here).
One explanation is that medium and low risk vulnerabilities are considered `lower hanging fruit' because there are several types of vulnerabilities and a moderate effort is likely to be enough to fix at least some of them.
(The severity level is also considered high according to OWASP.)
Therefore, students seem to be motivated enough to address these risks in an attempt to score some bonus points. 
There is a significant number of informational risks ($\approx$26\% of total vulnerabilities in all the runs) but since this type of risk is not considered severe, students did not seem to pay much attention.

\begin{table}[!t]
\caption{Percentage of vulnerability risk levels in last two runs' projects vs.~this year's (with and without the outlier)}
\label{tab:risk}
	\centering\footnotesize
	\begin{tabular}{lccc}
		\hline
		Risk Level & last 2 runs & this year & this year  \\ 
		& & w. outlier & w.o outlier \\ \hline
	    High & 0.52 & 0.35 & 0.49 \\
	    Medium & 12.75 & 22.26 & 10.59 \\
	    Low & 61.08 & 51.54 & 52.72 \\
	    Informational & 25.65 & 25.85 & 36.2
	\end{tabular}
\end{table}

\begin{table}[!t]
\caption{Percentage of types of vulnerabilities in last two runs' projects vs.~this year's (with and without the outlier)}
\label{tab:vulntype}
	\centering\footnotesize
	\begin{tabular}{lccc}
		\hline
		Type & last 2 runs & this year & this year  \\ 
		& & w. outlier & w.o outlier \\ \hline
	    Broken Access Control & 41.3 & 48 & 47.4 \\
	    Injection & 1.68 & 0.37 & 0.52 \\
	    Insecure Design & 1.04 & 0.18 & 0.26 \\
	    Security Misconfiguration & 41.86 & 40.1 & 35.8 \\
	    Integrity issue & 14.14 & 11.3 & 16
	\end{tabular}
\end{table}

ZAP also reports types of the vulnerabilities.
To understand which types of vulnerabilities students addressed better in this year's run after our intervention, we also looked at the scan results with respect to vulnerability type (Table~\ref{tab:vulntype}).

\begin{figure}[!t]
\begin{center}
\includegraphics[scale=0.35]{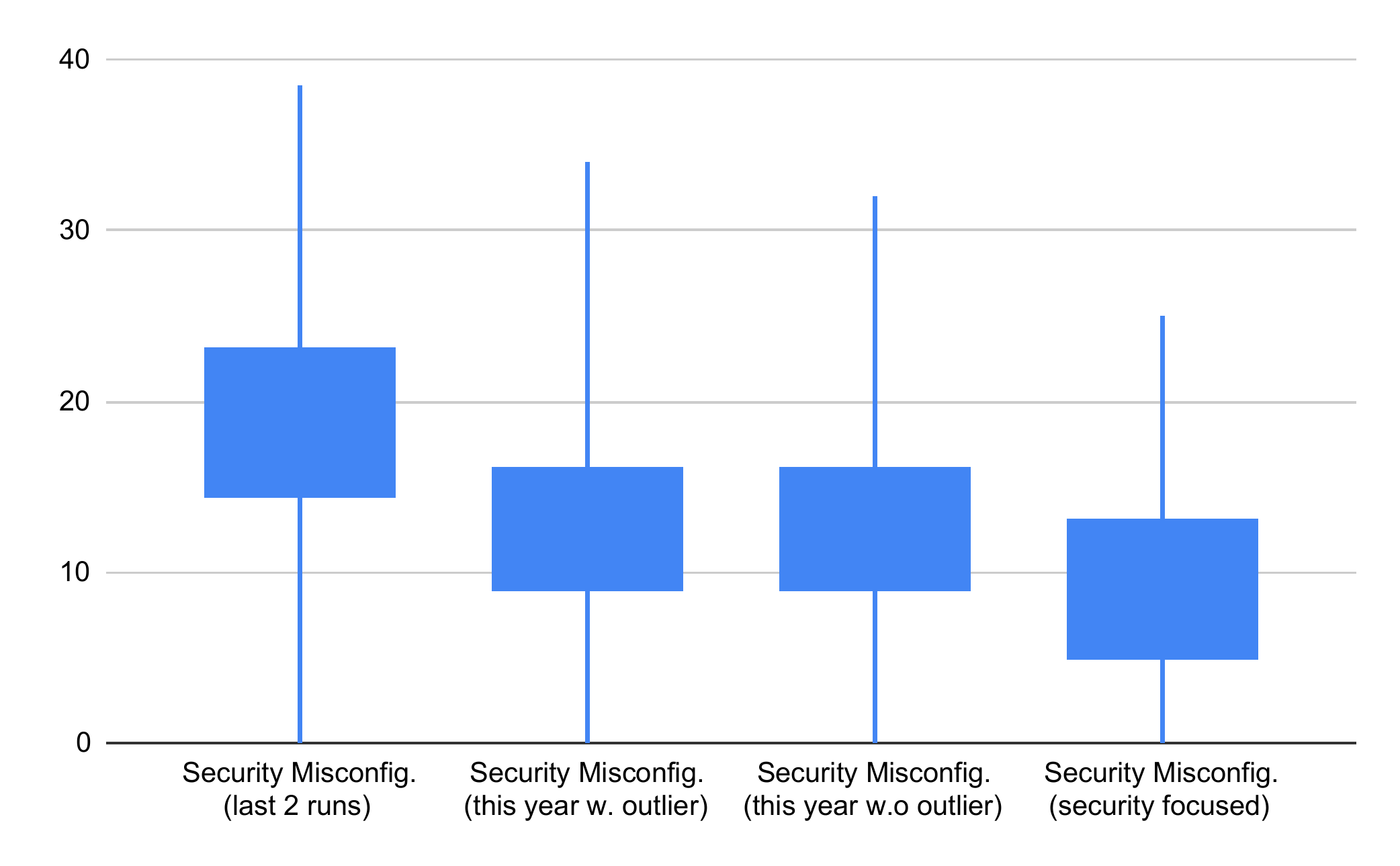}
\end{center}
\vspace{-1.5em}
\caption{Misconfiguration vulnerabilities found in this year's projects vs.~past years' vs.~security-focused projects}
\label{fig:misconfig}
\vspace{-1em}
\end{figure}

As shown in Table~\ref{tab:vulntype}, broken access control and security misconfiguration vulnerabilities are commonly found in students' projects.
After our intervention this year, we observed fewer security misconfiguration problems, including injection and insecure design problems, compared to the previous two runs. 
In terms of the distribution among the projects, the drop in security misconfiguration vulnerabilities is significant, as shown in the boxplots of Figure~\ref{fig:misconfig}.
One possible explanation of this is that security misconfiguration problems are generally easier to fix than others, e.g.~by using framework-provided security features.

The rightmost boxplots in Figure~\ref{fig:totalvuln}, Figure~\ref{fig:medium}, Figure~\ref{fig:low}, and Figure~\ref{fig:misconfig} show the results of the projects in this year's run which focused on security. 
We can observe a clear improvement in security based on these figures.
We inspected some of these projects' code to understand how they were able to achieve good security results.
We observed that, in these projects, security misconfiguration errors were mainly avoided by using framework-provided security features such as using the Firebase authentication service to provide secure authentication, and the Firebase environment to store API keys securely.
 
For example, the following listing shows the secure configuration method adopted by one of the teams that emphasised security:
\begin{lstlisting}[basicstyle=\footnotesize]
cred_dict = {
  "type": os.environ.get('FB_ACC_TYPE'),
  "private_key": os.environ.get('FB_PRIV_KEY').
                        replace('\\n', '\n'),
  "auth_uri": os.environ.get('FB_AUTH_URI'),
  "token_uri": os.environ.get('FB_TOKEN_URI'),
  "auth_provider_x509_cert_url": os.environ.get
                ('FB_AUTH_PROVIDER_CERT_URL')
}
\end{lstlisting}

The above code essentially stores the sensitive information (such as API keys) in environment variables and accesses them securely. In comparison, the following shows the insecure configuration (exposure of an API key in client code) used in a past year's project:

\begin{lstlisting}[basicstyle=\footnotesize]
const key="i9IigYi6bl70KMqOcpewpzHHQ2NanEqx";
var base_url="...";
var url=base_url + "?dataset="+dataset+ 
                            "&apikey="+key;
request.open("GET", url, true);
\end{lstlisting}

\section{Focus Groups \& Reflections}
In this section, we critically reflect on our experiences of introducing web security in an undergraduate web programming course.
To obtain some reflections from our students, we designed a short and optional survey which we distributed to them after the completion of all assessments and grading.
Our survey consisted of several `yes/no' questions (e.g. Did you attend the Week 8 web security talk? Did you implement web security in your project? Did you have prior experience/knowledge in web security?).
Our survey also consisted of Likert scale questions (e.g. How relevant is web security to front-end web development? Will you want to learn more about web security in the future? Should web security be part of web application development curricula?), and we asked students to indicate the level of their agreement using a five-point scale.
Finally, we asked open-ended questions about the web security tools they used in their project and what security vulnerabilities they were successful at resolving.
We collected 147 responses from the seven sections surveyed (managed by two faculty members), i.e.~a response rate of approximately 33\%.
Furthermore, we dived deeper into the students' perceptions by conducting two focus group sessions with four volunteers from the survey respondents.

\textbf{`Security' is perceived to be a hard, technical, advanced, daunting topic}: While about 15\% of the survey respondents indicated that `security' is a foreign topic to approach as part of their web programming learning, a focus group session revealed more in-depth insights into why this was the case.

\say{\textit{Having learnt security in polytechnic, I realised you cannot jump into security right away. You need some basic knowledge of computer networking first.}}

Students with prior exposure to security voiced that security was not an independent subject but a broad field of study that typically started with an introductory course in computer networking.
Currently, our information systems undergraduate curriculum does not offer a computer networking course.
Hence, when the subject of security came up in the web programming course in the middle of the term, many students found the subject to be daunting.

To make it less daunting for sophomores taking a web development course, short bridging courses during school breaks could be helpful.
A bridging course on computer networking during the summer break after the freshman year can provide students with sufficient knowledge of how digital systems are connected and where vulnerabilities could occur.

\textbf{The core course content is already overwhelming. Adding security as another topic will make the course very packed}: 39\% of the survey respondents cited `time constraints' to be a major factor behind not implementing web security in their projects.

\say {\textit{Only I had prior background in security. Though other front-end related bonuses were also new to us, my group members decided to do front-end bonuses such as Sass and visualisation libraries [instead of security]. It is more directly related to the core content we learnt in the course. Doing security would require more research, and we had little time left to finish the project.}}

By the sixth week of the term, most of the CSS and JavaScript concepts were taught sufficiently.
By then, most student groups start brainstorming on project details, including bonuses.
In selecting bonuses, many students did not find `web security' to be an \textit{approachable} topic.
Such groups instead pursued CSS and JavaScript-related frameworks and libraries.

\say{\textit{(Referring to Sass and animation libraries) These seemed more approachable than security because it's an extension of what we already learnt in class. Approachable... but then these came with [their] own learning curve and resulted in many hours of researching and experimenting.}}

It is worthwhile to note that information systems sophomores take on three core courses concurrently in the Fall term on top of one or two other track courses.
Given that about nine hours worth of work is expected for each university course, our students spend between 36 and 45 hours weekly on course work.
In each course, a group project (consisting of 5--6 students) is a substantial grade component, typically ranging between 25\% and 45\%.

\say{\textit{I was the only one in my group with some prior experience in security. I was keen to explore web security in our project, but my group mates voted against it because it was too risky compared to other more approachable bonuses.}}

Depending on the group composition and their willingness to go above and beyond, the group project component can introduce additional workload on top of the core course work.
Given the `time constraint', many groups chose \textit{safer} options and `security' was not one of them.

The aforementioned `bridging' courses that students can take on during school breaks can help reduce the amount of independent research students would have to perform if they were to implement security into their web development projects.

\say{\textit{Every year, seniors and school clubs offer technical workshops focused on application development, data structures and algorithms, etc. Workshops focused on security are hard to come by. Faculty-led bridging courses will be definitely helpful, but ideally, students and school clubs too will initiate offering workshops on security.}}

\textbf{Doing `security' later is challenging because it requires a re-design of the web application}: Our intervention of introducing `web security' in Week 8, right in the middle of the term, was met with mixed responses.
While some students indicated that they found the topic to be interesting and helpful, others indicated that the topic could have been introduced earlier.

\say{\textit{After developing a web app without security in mind, going back to implement web security was really challenging because we had to roll back our code changes and coordinate across multiple group members... every small change I made to implement web security resulted in more bugs for other group members}}

Those groups that implemented web security voiced that they did so after much of their web application had been already implemented (around Week 10).
In those groups, typically, only one group member took charge of the `web security' implementation.
Upon making code changes to implement web security, they found out that those changes led to other parts of their application breaking.
This resulted in many hours spent on fixing the `new' bugs and caused much frustration amongst the group members.

\say{\textit{(With prior experience in security) Very challenging to revert back and implement security. To implement security into web app, group must do the thinking right from the start.}}

In other words: secure coding practices need to be introduced as early as possible in the course.

\say{\textit{Instead of learning security as yet another topic in [the] web app dev course, where and when possible during lessons, faculty can highlight good and bad coding/design practices.}}

For example, when students learn about AJAX API calls, faculty can show examples of security vulnerabilities associated with them.
Instead of showing one `right' way to make API calls, faculty can show how ill-intended users can exploit them and how to protect web applications from such exploitation.

\textbf{Is `security' really an essential topic for all Information Systems students?}
When asked ``Should web security be part of web application development curricula?'', 71\% of the survey respondents indicated `Probably yes' or `Definitely yes'.
A question was raised by some students, and it pertains to whether `security' is really an essential topic in information systems curricula.

\say{\textit{I aspire to become a data analyst. I don't intend to become a system architect or application developer. What I learned in the current web programming course was already very helpful. With what I learned in the course, I would be able to build interactive data analysis dashboards. I'm not sure why `security' is so important for this purpose. I'd rather spend more time learning about data-related things such as data APIs and how to retrieve data from external APIs.}}

Currently, the percentage of students wanting to become software professionals upon graduation is far less compared to students aspiring to enter non-software job roles such as data analysts.
`Security' is perceived to be a \textit{niche} area amongst our students.
Those students not intending to pursue security-related or software-related job roles do not consider (rightly or wrongly) `security' to be an important and essential topic in their information systems learning.

This perception, however, contrasts with the fact that digital devices are becoming more predominant in daily life across the globe, including in the lives of younger children, potentially exposing them to fraudsters and hackers.
Thus, `security' awareness education should occur much earlier than university so that children are equipped with basic knowledge of risks associated with the digital devices they use~\cite{Mee20a}.
The students currently questioning whether security is an essential topic in their information systems learning lacked early education in security.
A greater emphasis on security and risks involved in using digital systems could be incorporated into an existing freshman course focusing on digital transformation.
When students are aware of the risks and the myriad of ways that they can be exploited, they will be motivated to learn how to protect digital systems against cyber attacks.

For the survey question ``Will you want to learn more about web security in the future?'', 89\% of the survey respondents indicated `Probably yes' or `Definitely yes'.
And again, 71\% indicated `Probably yes' or `Definitely yes' to the question ``Should web security be part of web application development curricula?''
A majority of the students seem to recognise the relevance and importance of security in IS216.
With the suggested bridging courses, and covering motivational real-life cases of cyber attacks in their freshman year, students embarking on web development courses in the sophomore year will be able to attain hands-on knowledge and skills to build secure web applications.

\textbf{Instructors' Reflections}: Of the four instructors that taught the web development course, only two instructors were knowledgeable about web security or had practical experience of implementing it in the past.

\say{\textit{I've knowledge of basic security concepts in the web context such as SQL injection and cross-site scripting (XSS), more from the perspective of what are the implications of each. Not having had practical implementation experience, I realised I would need some training in this area if I were to teach it to my students. Or perhaps, I am no longer a suitable instructor for this course if web security were to become a big part of the course.}}

Changing the web programming course curriculum would require an evaluation of the capabilities of the current course instructors.
The curriculum committee will have to investigate and suggest what kinds of security knowledge and skills are required from the course instructors, whether it is attainable through training, and if so, how much effort would be required.

\section{Conclusion}

In this experience report, we described our implementation of the security integration approach for an undergraduate web programming course.
In particular, we added a practical web security lecture to the course, which highlighted the OWASP Top 10 vulnerabilities by example, and showed students how to identify many of them in their code using out-of-the-box security scanners (e.g.~ZAP).
Furthermore, we incentivised students to integrate secure coding practices and the use of security scanners in their group projects by offering bonus marks to those who did.

We assessed our intervention by scanning three years of student projects, finding a lower distribution of vulnerabilities in the most recent run of the course.
Furthermore, in focus groups and a survey, we found that our intervention helped to raise general security awareness among students.
At the same time, we learnt that several students did not integrate secure coding in their projects because the lecture took place after they had already started coding, and the grading incentives were not high enough relative to the perceived effort.
This suggests to us that we are on the right path, but need to be bolder in introducing security material earlier (e.g.~interleaved throughout the regular JavaScript classes, or in summer enrichment courses), and could perhaps consider making secure coding a core part of the project rubrics.

In future iterations of the course, we would also be keen to explore the use of static analysis tools for finding potential security issues, e.g.~Semgrep~\cite{Semgrep}, which may be more effective at finding more challenging XSS vulnerabilities.
Unlike ZAP, however, static analysis tools return false positives, and it may be challenging to instruct students how to identify these and filter them out.

\bibliographystyle{ACM-Reference-Format}
\balance
\bibliography{references}

\end{document}